\pdfoutput=1
\documentclass[11pt,a4paper]{article}

\usepackage[margin=1in]{geometry}
\usepackage{amsmath,amssymb}
\usepackage{booktabs}
\usepackage{graphicx}
\usepackage{microtype}
\usepackage{caption}
\usepackage{xcolor}
\usepackage[numbers,sort&compress]{natbib}
\usepackage[colorlinks=true,linkcolor=blue!55!black,citecolor=blue!55!black,
            urlcolor=blue!55!black]{hyperref}
\hypersetup{pdftitle={Complementary-gap post-selection in the surface code under depolarizing noise},
            pdfauthor={Tushar Pandey}}
\usepackage{orcidlink}

\newcommand{\LA}{L_{\mathrm{A}}}
\newcommand{\LB}{L_{\mathrm{B}}}

\title{\bfseries Complementary-gap post-selection in the surface code under depolarizing noise}

\author{Tushar Pandey\,\orcidlink{0000-0001-7448-5723}}

\date{November 2026}

\begin{document}
\maketitle

\begin{abstract}
\noindent
Optimal post-selection ranks syndromes by the posterior of the most likely class, while matching
decoders rank them by the complementary gap. We show that for the rotated surface code under
code-capacity depolarizing noise, the gap of independent X and Z matching at distance 11 and half
acceptance leaves a failure rate $310$ times the optimum at $p=0.08$, $51$ times at $p=0.11$ and about
$1300$ times at $p=0.05$, against $5.4$ under uncorrelated X and Z noise with the marginal rates of
$p=0.08$. We compute every class posterior by boundary matrix-product-state contraction of four
stabilizer cosets. Ranked by this posterior, the matching decision is within $1.2\%$ of optimal up to
half acceptance, so the loss lies in ranking. Reweighting for Y errors removes about half of the logarithmic loss, but its gap leaves a factor
$10$ to $23$ averaged over acceptance, against about $3$ under uncorrelated noise. At $p=0.08$ the
acceptance-averaged loss of the independent gap grows with distance up to $d=13$.
\end{abstract}

\section{Introduction}

The complementary gap of minimum-weight perfect matching (MWPM), the difference between the weight of
the best correction and the weight of the best correction in a different logical class, is the standard
measure of decoder confidence, obtained from two constrained matchings \cite{meister2024,gidney2023yoked}.
Fault-tolerant protocols use such confidence to accept or reject the output of a decoding round:
low-overhead magic-state preparation and cultivation discard runs whose syndromes are ambiguous
\cite{bombin2022postselection,gidney2024cultivation}, yoked surface codes use it to flag unreliable inner
patches for the outer code \cite{gidney2023yoked}, and exclusive decoders reject difficult instances to
suppress logical errors \cite{smith2024}. Related soft outputs include the use of soft measurement
information \cite{pattison2021} and forced-gap post-selection for quantum low-density parity-check codes
\cite{wills2026}.

The optimal acceptance rule is known. For a fixed acceptance fraction, the lowest post-selected failure
rate is obtained by ranking syndromes by the posterior probability of the most likely logical class,
the maximum-likelihood (ML) class. This is Chow's optimal reject rule \cite{chow1970}, derived for
quantum error correction in Ref.~\cite{english2024}. Exact evaluation of that posterior is a
partition-function computation, which tensor-network and Monte Carlo decoders approximate to high
accuracy for the surface code under depolarizing, biased and non-Pauli noise
\cite{bravyi2014,hutter2014,darmawan2017,darmawan2018,tuckett2018,tuckett2019}. Maximum-likelihood post-selection has been studied for the toric code with perfect syndrome
measurements under independent bit- and phase-flip noise, where it raises the logical failure rate to a
power $b=3.1(1)$ \cite{chen2025}, and, with a neural-network estimate of the posterior, for the surface
code under code-capacity depolarizing noise, where a soft rule that treats the uncertainties of the two
logical operators separately was proposed \cite{seip2026}. For matching, a learned confidence
outperformed the complementary gap at fixed acceptance under circuit-level noise \cite{dentelski2026}.
None of these works compares the complementary gap with the exact posterior for noise that correlates
the X and Z parts of an error. Depolarizing noise is the simplest such case: a Y
error is a coincident X and Z error, a correlation that matching applied to the two parts independently
does not use, and that correlation-aware decoders exploit
\cite{fowler2013,delfosse2014,criger2018,nickerson2019,higgott2023belief}. Its effect on optimal
decoding is captured by the statistical-mechanical models of correlated noise \cite{chubb2021statmech}.

We compute the class posterior of every sampled syndrome of the rotated surface code under
code-capacity depolarizing noise by a boundary matrix-product-state (MPS) contraction of the four
stabilizer cosets \cite{bravyi2014,chubb2021}, converged at bond dimension 16 up to distance 11 through
exact application and canonical compression of each row. We show that the complementary gap of
independent matching loses to the optimal rule by orders of magnitude in post-selected failure rate, and
that the loss lies in the ranking of syndromes, not in the decision on the accepted ones. Two controls,
independent X noise and independent X and Z noise at the same marginal rates, compare the loss under
depolarizing noise with that caused by the gap's discreteness and by the two-part combination alone. We
further show that the complementary gap of correlation-aware matching, with edge weights reweighted for
Y errors, recovers about half of the logarithmic loss. The remaining loss grows with distance at
$p=0.05$ and $0.08$, and at $p=0.08$ its step from $d=11$ to $13$ is the smallest of the series.

Section~\ref{sec:model} defines the model and the four acceptance rules, and
Section~\ref{sec:method} the coset contraction and its validation. Section~\ref{sec:results} gives the
post-selected failure rates and Section~\ref{sec:origin} traces the loss to the ranking.
Section~\ref{sec:ca} treats correlation-aware matching, Section~\ref{sec:scaling} the dependence on
distance, and Section~\ref{sec:discussion} discusses the mechanism and the scope of the comparison.

\section{Model and acceptance rules}
\label{sec:model}

The code is the rotated surface code of odd distance $d$ with $d^2$ data qubits. Under code-capacity
depolarizing noise each qubit independently suffers X, Y or Z with probability $p/3$ each. An error is
written $(e_x,e_z)\in\{0,1\}^{2d^2}$, with $e_x$ detected by the Z-type faces and $e_z$ by the X-type
faces. Its logical class is $k=(k_x,k_z)$ with $(k_x,k_z)=(\ell_Z\cdot e_x,\,\ell_X\cdot e_z)\bmod 2$,
where $\ell_Z$ and $\ell_X$ are the supports of the logical Z and X operators (a column and a row of the
lattice). Given the syndrome $s$ and any error $(x_0,z_0)$ that produces it, the probability of class
$k$ is the coset sum
\begin{equation}
Z_k(s) = \sum_{g} \prod_{q=1}^{d^2}
w\bigl(x_{0,q}\oplus a\,\ell_{X,q}\oplus (S_X g)_q,\; z_{0,q}\oplus b\,\ell_{Z,q}\oplus (S_Z g)_q\bigr),
\label{eq:coset}
\end{equation}
where $(a,b)$ is the class offset of $k$ relative to $(x_0,z_0)$, $g\in\{0,1\}^{d^2-1}$ selects a
subset of the stabilizer generators, so that the sum runs over the $2^{d^2-1}$ elements of the
stabilizer group, $S_X$ and $S_Z$ are the $d^2\times(d^2-1)$ matrices of the X and Z supports of the
generators (arithmetic mod 2), and
$w(0,0)=1-p$, $w(x,z)=p/3$ otherwise. The posterior is $P(k\mid s)=Z_k(s)/\sum_{k'}Z_{k'}(s)$, and the ML
decoder returns its maximizer $\hat k$. Its conditional failure probability is
$q_{\mathrm{ML}}(s)=1-P(\hat k\mid s)$.

The matching decoder applies MWPM separately to the X part at the marginal rate $2p/3$ and to the Z
part at the same rate, with edge weight $\ln[(1-2p/3)/(2p/3)]$ on every edge, boundary edges included,
using PyMatching \cite{dennis2002,higgott2023}. Its class $k_{\mathrm{M}}$ has conditional failure probability
$q_{\mathrm{M}}(s)=1-P(k_{\mathrm{M}}\mid s)$. The class-restricted minimum weights $W_0$, $W_1$ of each
part are obtained, as in Refs.~\cite{gidney2023yoked,meister2024}, by adding one detector node on the
support of the corresponding logical operator: the parity of that node in a correction is its logical
class, so setting the node's syndrome bit to $0$ or $1$ restricts the matching to one class. Since
$W_c$ is indexed by the logical parity $c$ of the correction, not relative to the matching class, the
complementary gap is $G_X=|W_1-W_0|$ for the X part, and $G_Z$ likewise for the Z part. The smallest weight increase that changes the joint class is
$G=\min(G_X,G_Z)$.

We compare four acceptance rules. Each accepts the fraction $\alpha$ of shots with the highest score
and keeps the stated decision:
\begin{center}
\begin{tabular}{lll}
\toprule
rule & score & decision \\
\midrule
A & $P(\hat k\mid s)$ & ML \\
B & $G=\min(G_X,G_Z)$ & matching \\
B2 & $\sigma(G_X)\,\sigma(G_Z)$, $\sigma(g)=1/(1+e^{-g})$ & matching \\
C & $P(k_{\mathrm{M}}\mid s)$ & matching \\
\bottomrule
\end{tabular}
\end{center}
Among rules that accept on the syndrome, rule A is optimal at every $\alpha$ \cite{chow1970,english2024}: the
post-selected failure rate is the mean of $q(s)$ over the accepted
syndromes, and at fixed acceptance probability this mean is smallest when the syndromes of lowest
$q_{\mathrm{ML}}(s)$ are accepted first, with random tie-breaking at the cut. Split rules of the kind of
Ref.~\cite{seip2026} can therefore match but not improve on rule A at equal acceptance. Rule B is the
standard gap rule. If each gap were the exact log-likelihood ratio of its part's two classes and the
parts were independent, $\sigma(G_X)\,\sigma(G_Z)$ would be the probability that both parts are decoded
correctly. Rule B2 is this soft output implied by independence. It orders shots approximately by $G$ and, within a gap value, by the gap
of the other part. Rule C keeps the matching decision but ranks by its
converged posterior, which separates the cost of the decision from the cost of the ranking:
$\LB/\LA=(L_{\mathrm{C}}/\LA)(\LB/L_{\mathrm{C}})$.

Because acceptance depends on the syndrome alone, the post-selected failure rate of a rule is the mean
of its conditional failure probability over the accepted shots, $L=\langle q\rangle_{\mathrm{acc}}$.
For a fixed acceptance region this estimator is unbiased. With the cut set at a sample quantile it is
consistent, with a bias of order $1/n$ for $n$ shots. It reaches rates far below $1/n$, which counting
failures cannot. At $\alpha=0.01$ and $n=10\,000$ the accepted set has 100 shots. Where more shots than
that have the empty syndrome, rules A and B accept empty syndromes only, which gives the flat segments
of the curves in Figure~\ref{fig:curves}. The gap takes discrete values (odd multiples of the edge weight), so rule B meets ties
at the acceptance cut. Every rule is therefore evaluated at its exact expectation over uniformly random
tie-breaking: the shots tied at the cut enter with their mean $q$, weighted by the number of places left.
Statistical errors are standard deviations over 300 bootstrap resamples of the shots, paired: the same
resample is used for every rule.

\section{Class posterior by coset contraction}
\label{sec:method}

Equation~\eqref{eq:coset} is contracted as a two-dimensional network with one binary variable per
stabilizer face. The face variables of one row of faces form an MPS with one site per face column.
Each row of qubits couples two consecutive rows of faces. The corresponding factors are nonnegative
$4\times4$ matrices over the old and new face bits of two neighboring sites. A row
is applied exactly by splitting every factor into a rank-4 matrix-product operator, the old face bits
are summed, and the MPS is compressed to bond dimension $\chi$ by a left-to-right QR sweep followed by
a right-to-left truncated singular-value sweep, in which each bond is truncated optimally in the 2-norm
given the others \cite{schollwock2011}. Each of the four classes is its own contraction with
nonnegative tensors, so no class is formed as a difference of larger quantities. The resulting relative
accuracy of each class is checked numerically in Table~\ref{tab:validation}. The reference error
$(x_0,z_0)$ is the matching correction, and class $(0,0)$ is the matching class. The posterior at
$\chi=16$ is referred to as converged below. Its truncation error is given in
Table~\ref{tab:validation}.

\begin{table}[t]
\centering
\caption{Validation of the contraction and of the gap computation.}
\label{tab:validation}
\footnotesize
\begin{tabular}{lll}
\toprule
check & conditions & result \\
\midrule
class sums vs.\ sum over $4^9$ errors & $d=3$, 256 syndromes, $p\in\{0.05,0.15\}$ & max.\ rel.\ error $2.8\times10^{-13}$ \\
posterior, $\chi=16$ vs.\ $128$ & $d=9$, 32 shots, $p\in\{0.12,0.17\}$ & zero difference (no truncation) \\
class sums, $\chi=16$ vs.\ $128$ & $d=11$, 32 shots, $p\in\{0.12,0.17\}$ & max.\ rel.\ error $1.3\times10^{-4}$ \\
posterior, $\chi=16$ vs.\ $128$ & $d=11$, 32 shots, $p\in\{0.12,0.17\}$ & max.\ abs.\ error $8.2\times10^{-6}$ \\
posterior, $\chi=32$ vs.\ $128$ & $d=11$, 32 shots, $p\in\{0.12,0.17\}$ & zero difference (no truncation) \\
$q_{\mathrm{ML}}\ge9.7\times10^{-11}$, $\chi=16$ vs.\ $32$ & $d=11$, $p=0.08$, 100 shots & max.\ rel.\ error $3.7\times10^{-9}$ \\
posterior, $\chi=16$ vs.\ $32$ & $d=11$, $p=0.05$, 10\,000 shots & max.\ abs.\ error $1.1\times10^{-11}$ \\
$q_{\mathrm{ML}}\ge7.8\times10^{-16}$, $\chi=16$ vs.\ $32$ & $d=11$, $p=0.05$, 10\,000 shots & max.\ rel.\ error $6.3\times10^{-7}$ \\
class-restricted weights & $d=3$, all syndromes, both parts & max.\ abs.\ error $8.9\times10^{-16}$ \\
posterior, $\chi=16$ vs.\ $32$ & $d=13$, $p=0.08$, 140 shots & max.\ abs.\ error $7.2\times10^{-10}$ \\
X-noise posterior vs.\ X decoder & $d\in\{5,7\}$, 120 shots & max.\ abs.\ error $2.2\times10^{-14}$ \\
\bottomrule
\end{tabular}
\end{table}

Table~\ref{tab:validation} lists the checks. At $d=11$ the canonical compression reaches a posterior
error of $8.2\times10^{-6}$ at $\chi=16$ in $0.43$\,s per shot on one core, and $\chi=32$ agrees with
$\chi=128$ exactly on the same shots, so no truncation occurs there. We recompute the first $10\,000$ shots of the $p=0.05$, $d=11$ cell
at $\chi=32$: the class posteriors differ from those at $\chi=16$ by at most
$1.1\times10^{-11}$ (mean $2.6\times10^{-15}$), no ML decision changes, and $q_{\mathrm{ML}}$ agrees to a
relative $6.3\times10^{-7}$ down to $q_{\mathrm{ML}}=7.8\times10^{-16}$. We run all production cells at
$\chi=16$ with 14 worker processes. For the X-noise control the single-qubit weights are $(1-p_x,0,p_x,0)$ in the order
$(\mathrm{I},\mathrm{Z},\mathrm{X},\mathrm{Y})$, the Z-part classes are omitted and the Z-part gap is
infinite.

\section{Post-selected failure rates}
\label{sec:results}

We draw shots at $p\in\{0.05,0.08,0.11\}$ and $d\in\{5,7,9,11\}$: 10\,000 shots per distance at
$p=0.05$ and $0.11$, and 20\,000 at $p=0.08$ (18\,775 at $d=5$, from two runs), with a further
100\,000 shots from a second seed at $d=11$, $p=0.05$. All three values lie
below the threshold of independent matching, about $0.155$ (the X-noise matching threshold of about
$0.103$ \cite{dennis2002,wang2003} applied to the marginal rate $2p/3$, see also \cite{english2024}),
and below the ML threshold of about $0.189$ \cite{bombin2012,chubb2021statmech}. The first control uses
independent X noise at $p_x=0.0533$, the X-part rate of depolarizing noise at $p=0.08$, with 10\,000
shots per distance (10\,280 at $d=5$, including a separately seeded 280-shot validation run). The
second control uses independent X and Z errors, each at rate $m=0.0533$, with 10\,000 shots per
distance: its two parts have the same marginal statistics as those of depolarizing noise at $p=0.08$
and rule B keeps the form $\min(G_X,G_Z)$, but the parts are uncorrelated and matching at rate $m$ is
the exact minimum-weight rule for each part. The single-qubit weights are then
$\bigl((1-m)^2,m(1-m),m(1-m),m^2\bigr)$, and failure probabilities are summed over the non-chosen
classes, which keeps them accurate below $10^{-13}$.

Without post-selection the two decoders differ by the ML advantage alone. At $d=11$ the failure rates
are $3.11\times10^{-4}$ (ML) and $1.69\times10^{-3}$ (matching) at $p=0.05$, $4.67\times10^{-3}$ and
$1.91\times10^{-2}$ at $p=0.08$, and $2.76\times10^{-2}$ and $8.17\times10^{-2}$ at $p=0.11$. Under
X noise they are $9.54\times10^{-3}$ and $9.93\times10^{-3}$, and under independent X and Z noise
$1.84\times10^{-2}$ and $1.92\times10^{-2}$.

\begin{figure}[t]
\centering
\includegraphics[width=\linewidth]{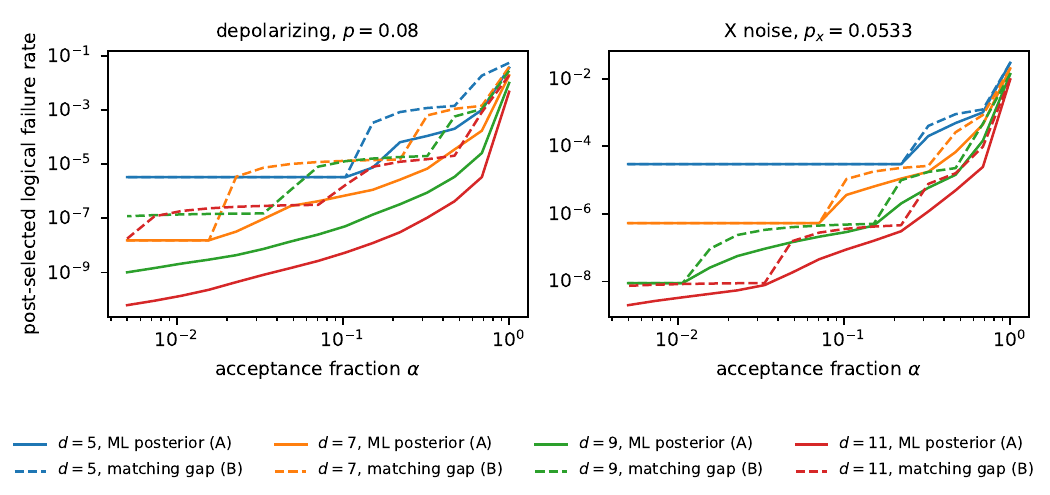}
\caption{Post-selected logical failure rate against acceptance fraction for the optimal rule A (solid)
and the gap rule B (dashed). Left: depolarizing noise at $p=0.08$. Right: independent X noise at
$p_x=0.0533$, the same X-part rate. Where the acceptance falls below the probability of the empty
syndrome, both rules accept the same shots and the curves coincide. The flat segments at small
$\alpha$, such as that of rule A at $d=5$, are this effect.}
\label{fig:curves}
\end{figure}

Figure~\ref{fig:curves} shows the post-selected rates as functions of $\alpha$, and
Table~\ref{tab:main} gives them at $d=11$. The acceptance fractions examined are
$\alpha\in\{0.01,0.02,0.05,0.1,0.2,0.5,1\}$ and, for averages over $\alpha$, nine log-spaced values in
$[0.02,0.5]$. Under depolarizing noise at $d=11$ the gap rule stays at least a factor $40$ above the
optimum at every one of these fractions between $0.01$ and $0.5$. The minimum, $44$ ($\ln=3.78$), occurs
at $p=0.11$, and the minima at $p=0.05$ and $0.08$ are $51$ and $128$. At full acceptance only the
decisions differ, and at smaller $d$ and small $\alpha$ both rules can accept only the empty syndrome,
where they coincide. At $\alpha=0.5$ the ratio $\LB/\LA$ is $310$ ($\ln=5.74\pm0.18$) at $p=0.08$ and
$51$ ($3.94\pm0.10$) at $p=0.11$. Under independent X and Z noise it is $5.4$ ($1.69\pm0.02$), and under
X noise $2.8$ ($1.02\pm0.03$). At $p=0.05$ it is $1260$ ($7.14\pm0.18$) on $110\,000$ shots, with
$\LA=2.3\times10^{-10}$ an average of failure probabilities down to $7\times10^{-16}$. Recomputed at $\chi=32$
on the first $10\,000$ shots, $\LA$ and $\LB$ of that subset change by relative amounts below $4\times10^{-9}$ at every acceptance fraction examined, and every
logarithmic ratio is unchanged to four decimals (Table~\ref{tab:validation}). The gap rule's rate at $p=0.05$ is also dominated by a few shots: the
five largest of the 110\,000 contributions carry $34\%$ of $\LB$. Deleting those five shots lowers the
logarithm to $6.72$, and the $2.5$ to $97.5\%$ bootstrap percentile interval is $[6.79,7.46]$, so the
factor is at least about $800$ in every such variation. At $p=0.08$ and $0.11$ the five largest
contributions carry $22\%$ and $14\%$. At $d=11$ and $\alpha=0.5$, ordering within a gap value by the
other part's gap (rule B2) recovers a factor between $1.3$ and $3$ of the depolarizing loss and leaves
the rest. Ranked by its posterior, the optimal rule reaches the failure rate of the gap rule at
$\alpha=0.5$ while accepting $95\%$, $94\%$ and $90\%$ of the shots at $p=0.05$, $0.08$ and $0.11$ (at
$p=0.08$ on the $40\,000$-shot cell of Section~\ref{sec:scaling}), and that of the correlation-aware gap of
Section~\ref{sec:ca} while accepting $81\%$, $80\%$ and $77\%$.

\begin{table}[t]
\centering
\caption{Post-selected failure rates at $d=11$ for the optimal rule ($\LA$) and the gap rule ($\LB$),
with the log-ratios of rules B, B2 and C to rule A. An entry $0$ is exactly zero (rules A and C accept
the same shots with the same decision). $^\dagger$The five largest shot contributions carry more than a
third of $\LB$ ($34\%$ in both cases), so the bootstrap standard deviation understates the
uncertainty. The $2.5$ to $97.5\%$ bootstrap percentile intervals of $\ln(\LB/\LA)$ are $[6.79,7.46]$
and $[3.62,5.94]$. The $p=0.05$ entries use $110\,000$ shots. On the first $10\,000$ of them a
recomputation at $\chi=32$ leaves the rates unchanged (Section~\ref{sec:method}).}
\label{tab:main}
\small
\setlength{\tabcolsep}{5pt}
\begin{tabular}{lcccccc}
\toprule
noise & $\alpha$ & $\LA$ & $\LB$ & $\ln(\LB/\LA)$ & $\ln(L_{\mathrm{B2}}/\LA)$ & $\ln(L_{\mathrm{C}}/\LA)$ \\
\midrule
depol., $p=0.05$ & 0.5 & $2.28\times10^{-10}$ & $2.87\times10^{-7}$ & $7.14\pm0.18^\dagger$ & $6.12\pm0.19$ & $0$ \\
                 & 0.1 & $1.07\times10^{-12}$ & $2.17\times10^{-9}$ & $7.61\pm0.13$ & $5.98\pm0.15$ & $0$ \\
depol., $p=0.08$ & 0.5 & $5.74\times10^{-7}$ & $1.78\times10^{-4}$ & $5.74\pm0.18$ & $4.63\pm0.18$ & $0.002\pm0.001$ \\
                 & 0.1 & $5.09\times10^{-9}$ & $1.04\times10^{-6}$ & $5.32\pm0.57^\dagger$ & $4.32\pm0.43$ & $0$ \\
depol., $p=0.11$ & 0.5 & $1.00\times10^{-4}$ & $5.15\times10^{-3}$ & $3.94\pm0.10$ & $3.69\pm0.11$ & $0.011\pm0.003$ \\
                 & 0.1 & $1.48\times10^{-6}$ & $1.37\times10^{-4}$ & $4.52\pm0.15$ & $4.04\pm0.15$ & $0$ \\
X and Z, $m=0.0533$ & 0.5 & $7.14\times10^{-5}$ & $3.88\times10^{-4}$ & $1.69\pm0.02$ & $1.50\pm0.04$ & $0$ \\
                  & 0.1 & $2.02\times10^{-6}$ & $1.71\times10^{-5}$ & $2.14\pm0.03$ & $1.75\pm0.06$ & $0$ \\
X, $p_x=0.0533$  & 0.5 & $6.14\times10^{-6}$ & $1.70\times10^{-5}$ & $1.02\pm0.03$ & $1.02\pm0.03$ & $0$ \\
                 & 0.1 & $8.36\times10^{-8}$ & $3.56\times10^{-7}$ & $1.45\pm0.04$ & $1.45\pm0.04$ & $0$ \\
\bottomrule
\end{tabular}
\end{table}

The rates of order $10^{-12}$ and below are expectations of converged conditional failure probabilities
under the noise model, not counted failures.\footnote{In the depolarizing data the failure probability
was stored as $1-\max_k P(k\mid s)$, which loses precision below about $10^{-13}$. This affects 132
of the first $10\,000$ shots at $p=0.05$, $d=11$ and none elsewhere. Raising those values to $10^{-13}$
changes $\ln(\LB/\LA)$ at $\alpha=0.5$ by less than $10^{-4}$ and the spread of Section~\ref{sec:origin}
by less than $0.001$ decades. The further $100\,000$ shots at $p=0.05$, $d=11$ store the sum of the
posteriors of the non-chosen classes, which keeps full precision.}

\section{Origin of the loss}
\label{sec:origin}

The last column of Table~\ref{tab:main} compares rule C with rule A. When the shots are ranked by the
converged posterior, keeping the matching decision instead of the ML decision costs almost nothing:
over all distances $d=5$ to $11$ and all noise models, $\ln(L_{\mathrm{C}}/\LA)$ is at most $0.011$ at
$\alpha=0.5$ and zero at $\alpha=0.1$.
The shots with the lowest failure probability are those on which both decoders agree. The loss of rule
B is therefore not a loss of decoding accuracy but of ranking: the gap accepts shots of high failure
probability and rejects shots of low failure probability.

\begin{figure}[t]
\centering
\includegraphics[width=\linewidth]{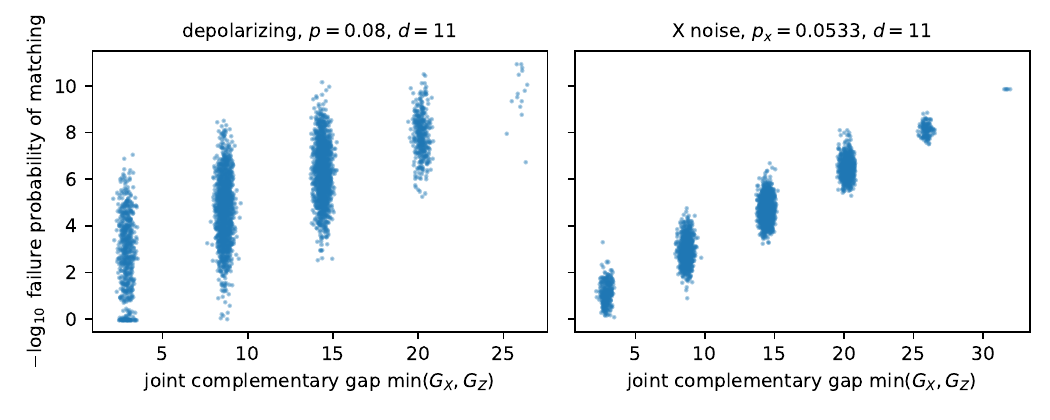}
\caption{Converged failure probability of the matching decision, plotted as $-\log_{10}q_{\mathrm{M}}$,
against its joint complementary gap for 4000 shots at $d=11$. The gap takes the odd multiples
$2.9$, $8.6$, $14.4$, $20.1,\dots$ of the edge weight $2.88$. Values are jittered horizontally for
visibility. Left: depolarizing noise at $p=0.08$. Right: independent X noise at the same X-part rate.}
\label{fig:calib}
\end{figure}

Figure~\ref{fig:calib} shows how the gap ranks shots. Under X noise the failure probability of the matching
decision is ordered by the gap, with moderate spread within each gap value. Under depolarizing noise the
spread within one gap value is much larger. Averaged over gap values with at least 50 shots, the
10th-to-90th percentile width of $\log_{10}q_{\mathrm{M}}$ within one value of $G$ at $d=11$ is $3.42$,
$3.35$ and $3.13$ decades at $p=0.05$, $0.08$ and $0.11$, against $1.21$ decades under X noise. The
Spearman rank correlation between $G$ and $-\ln q_{\mathrm{M}}$ is $0.78$, $0.73$ and $0.70$ in the
three depolarizing cells and $0.92$ under X noise.

The two controls measure the loss caused by the gap's discreteness and by the two-part combination
without X--Z correlation. Under X noise there is one part and no correlation, and the loss at $d=11$,
$\alpha=0.5$ is $1.02$ in the logarithm. It is the cost of a discrete score that cannot order the shots
inside one gap value, and it depends on where the cut falls among the gap levels: at $d=11$ it varies
between $0.04$ and $2.13$ over the seven acceptance fractions of Section~\ref{sec:results}, and at
$\alpha=0.5$ it is $0.59$, $1.26$, $0.31$ and $1.02$ for $d=5$ to $11$. Under independent X and Z noise
the rule combines two parts through $\min(G_X,G_Z)$, and the loss rises to $1.69$. Averaged over the
nine log-spaced acceptance fractions in $[0.02,0.5]$ it is $1.18$, against $1.17$ under X noise. The
within-level spread is $1.23$ decades and the rank correlation $0.90$, both close to the X-noise
values. The split between the discreteness and two-part terms depends on $\alpha$: the two-part term is
$0.67$ at $\alpha=0.5$ but $0.01$ averaged over $\alpha$. Depolarizing noise with the same marginals
gives $5.74$ at $\alpha=0.5$ and $5.69$ averaged over $\alpha$, so its loss is $50$ to $100$ times larger
than that of the control. The controls match the marginal rates, not the distance to threshold, so this
difference combines the effect of the correlation with that of the distance to threshold. Without post-selection the ML failure
rate under independent X and Z noise is $1.84\times10^{-2}$ at $d=11$, four times the depolarizing
value, and the depolarizing loss itself depends on $p$. A control matched in unselected failure rate
would separate the two effects.

The gap approximates the posterior of the model that independent matching assumes, independent X and Z
errors at rate $2p/3$, and that model differs from depolarizing noise. Rule D separates the two: it ranks
shots by the converged maximum class posterior of the independent model and keeps that model's ML decision,
so it removes the gap approximation and keeps the model mismatch. On the $20\,000$ shots at $d=11$,
$p=0.08$, $\ln(L_{\mathrm{D}}/\LA)$ averaged over $\alpha$ is $3.68\pm0.19$ against $5.69\pm0.25$ for
rule B ($3.54\pm0.15$ against $5.74\pm0.16$ at $\alpha=0.5$). The decision of rule D differs from that of matching
on $0.36\%$ of shots, and keeping the matching or the ML decision instead leaves the rates unchanged for
$\alpha\le0.5$. Rule D lies closer to B than to A: the model mismatch accounts for $65\pm1.5\%$ of the
logarithmic loss and the gap approximation for the remaining third.

\section{Correlation-aware matching}
\label{sec:ca}

Matching that uses the Y correlation reweights the edges of one part according to the correction of the
other \cite{fowler2013,delfosse2014,nickerson2019}. Given the correction $c$ of the other part, a qubit
flips in this part with probability $1/2$ where $c$ acts, since an X error there is accompanied by Z
with conditional probability $(p/3)/(2p/3)$, and with probability $(p/3)/(1-2p/3)$ elsewhere. The edge
weights are the corresponding log-likelihood ratios, so edges where $c$ acts have weight exactly zero,
which is why the conditioned gap can be small. The decoder starts from independent matching of one
part, alternates conditioned decodes of the two parts until neither correction changes, runs both
orders and keeps the pair with fewer qubits on which either part acts. The alternation converges on at
least $99.8\%$ of shots in every cell. On the remaining shots the last iterate is kept. Its complementary gap $G'=\min(G_X',G_Z')$ is formed from the class-restricted weights of
each part on its graph conditioned on the other part's final correction. The class-restricted weights
agree with a brute-force minimum over all $2^9$ patterns at $d=3$ to $2\times10^{-15}$. Rule B$'$ ranks by
$G'$ and keeps the correlation-aware decision. Rule C$'$ keeps that decision but ranks by its converged
posterior. Both are evaluated on the same shots as rules A and B, 10\,000 per cell and 110\,000 at
$d=11$, $p=0.05$.

Without post-selection, correlation-aware matching lies between independent matching and ML: at $d=11$
its failure rates are $5.73\times10^{-4}$, $8.40\times10^{-3}$ and $4.31\times10^{-2}$ at $p=0.05$,
$0.08$ and $0.11$. Over all distances and rates, rule C$'$ matches rule A to
$\ln(L_{\mathrm{C}'}/\LA)\le10^{-4}$ for $\alpha\le0.5$, so the correlation-aware decision is again not the source of loss.

\begin{table}[t]
\centering
\caption{Logarithm of the post-selected failure rate relative to the optimal rule, for the gap of
independent matching (B) and of correlation-aware matching (B$'$). The $\alpha=0.5$ columns use 10\,000
shots per cell, the same for B and B$'$, with the bootstrap re-evaluated. The average for B uses all
shots of the cell (20\,000 at $p=0.08$). At $p=0.05$ all four entries use the same $110\,000$ shots. The average runs over nine log-spaced acceptance fractions in
$[0.02,0.5]$. The uncorrelated controls give $1.17$ (X) and $1.18$ (X and Z) at $d=11$.
$^\ddagger$Value on the 10\,000-shot subset. On all 20\,000 shots the value is $5.74\pm0.18$
(Table~\ref{tab:main}). $^\dagger$The five largest shot contributions carry more than a third of the
post-selected rate ($34\%$ for B, $91\%$ for B$'$), so the standard deviation understates the
uncertainty. We compute percentile intervals only for B (Table~\ref{tab:main}).}
\label{tab:ca}
\small
\begin{tabular}{lcccc}
\toprule
& \multicolumn{2}{c}{$\alpha=0.5$, $d=11$} & \multicolumn{2}{c}{average over $\alpha$, $d=11$} \\
$p$ & B & B$'$ & B & B$'$ \\
\midrule
0.05 & $7.14\pm0.19^\dagger$ & $3.67\pm0.13$ & $5.88\pm0.09$ & $3.13\pm0.10$ \\
0.08 & $5.86\pm0.21^\ddagger$ & $4.02\pm0.88^\dagger$ & $5.69\pm0.25$ & $3.08\pm0.22$ \\
0.11 & $3.94\pm0.11$ & $2.43\pm0.10$ & $5.07\pm0.10$ & $2.34\pm0.09$ \\
\bottomrule
\end{tabular}
\end{table}

\begin{figure}[t]
\centering
\includegraphics[width=\linewidth]{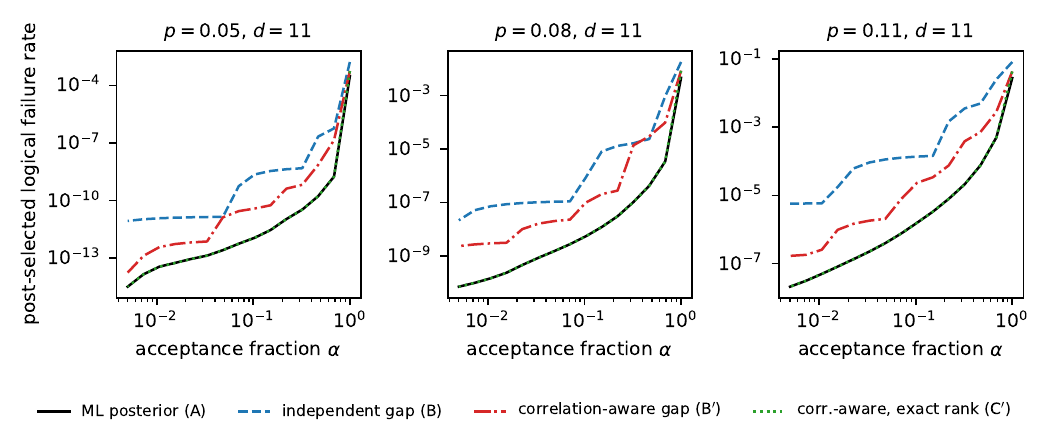}
\caption{Post-selected logical failure rate at $d=11$ against acceptance fraction for the optimal rule
(A), the gap of independent matching (B), the gap of correlation-aware matching (B$'$) and the
posterior ranking of the correlation-aware decision (C$'$, indistinguishable from A).}
\label{fig:ca}
\end{figure}

Table~\ref{tab:ca} and Figure~\ref{fig:ca} compare the two gaps. Correlation-aware reweighting reduces the
logarithm of the loss by about half: by $46$ to $54\%$ averaged over acceptance fractions and by $31$ to
$49\%$ at $\alpha=0.5$. Averaged over acceptance fractions at $d=11$, the remaining factor
is $e^{3.13}\approx23$, $e^{3.08}\approx22$ and $e^{2.34}\approx10$ at $p=0.05$, $0.08$ and $0.11$,
against about $3$ under the uncorrelated controls. At $p=0.08$ and $\alpha=0.5$ the five largest shot
contributions carry $91\%$ of $L_{\mathrm{B}'}$, so that entry is poorly determined. The average over
$\alpha$ is not affected in the same way. Within one value of $G'$ at $d=11$ the failure
probability of the correlation-aware decision spans $1.7$ decades between its 10th and 90th percentiles,
against $3.1$ to $3.4$ for the independent gap and $1.2$ for the uncorrelated controls, and the rank
correlation between $G'$ and $-\ln q$ is $0.90$ to $0.94$. Ordering within a gap value by the other
part's gap, as in rule B2, gives no consistent further gain: at $\alpha=0.5$ it lowers the logarithm by
$0.77$ at $p=0.05$ and by $0.08$ at $p=0.11$, and raises it by $0.06$ at $p=0.08$. The averaged loss of
rule B$'$ increases with distance at $p=0.05$ and $0.08$, from $0.27$ to $3.13$ and from $0.47$ to
$3.08$ between $d=5$ and $11$. At $p=0.08$ the steps are $0.92$, $1.04$ and $0.65\pm0.28$, so the
growth is slowing. At $p=0.11$ it rises from $0.62\pm0.02$ at $d=5$ to $2.29\pm0.13$ at $d=9$ and is
flat thereafter, $2.34\pm0.09$ at $d=11$.

The conditional gap holds one part fixed while the class of the other changes. Competing classes reached
by changing both parts at once, through chains of Y errors, are therefore still priced as two separate
changes. This interpretation is consistent with the residual spread and is a conjecture.

\section{Dependence on distance}
\label{sec:scaling}

\begin{figure}[t]
\centering
\includegraphics[width=0.55\linewidth]{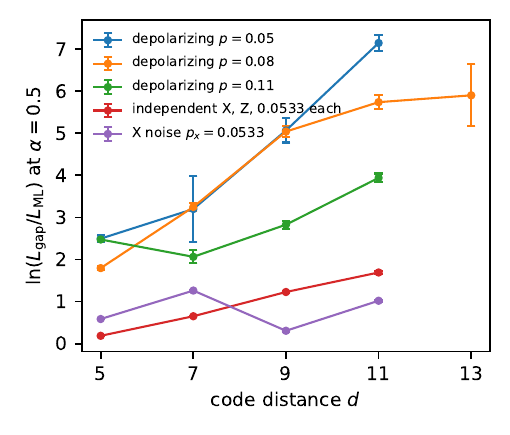}
\caption{Logarithm of the ratio of post-selected failure rates, gap rule to optimal rule, at
acceptance $\alpha=0.5$ against code distance, for depolarizing noise and for the two uncorrelated
controls. The $d=13$ point, at $p=0.08$ only, uses $40\,000$ shots. Error bars are bootstrap standard deviations.}
\label{fig:scaling}
\end{figure}

Figure~\ref{fig:scaling} shows $\ln(\LB/\LA)$ at $\alpha=0.5$ against $d$. At $p=0.05$ the points follow
a straight line with slope $0.76\pm0.03$ per unit of $d$ ($\chi^2=3.7$ on 2 degrees of freedom), so
each increase of the distance by two multiplies the loss by about $4.5$. At $p=0.08$ the weighted slope
is $0.73\pm0.02$, with $\chi^2=13.9$ on 2 degrees of freedom, so the straight line summarizes the trend
only roughly. The increase from $d=9$ to $d=11$ is $0.70$, against $1.79$ from $d=7$ to $d=9$. At
$p=0.11$ the loss at $\alpha=0.5$ grows from $d=7$ to $d=11$ but is not monotone from $d=5$. Because a
single $\alpha$ is sensitive to where the cut falls among the discrete gap levels, we also average the
logarithm over the nine log-spaced acceptance fractions in $[0.02,0.5]$. This average increases monotonically with
$d$ at every depolarizing rate: from $0.67$ to $5.88$ at $p=0.05$, from $1.17$ to $5.69$ at $p=0.08$ and
from $1.67$ to $5.07$ at $p=0.11$ between $d=5$ and $11$. At $p=0.05$ the $d=11$ cell holds $110\,000$
shots from two seeds. At $p=0.08$ we extend the shots of
Section~\ref{sec:ca} to $40\,000$ per distance (four seeds) at $d=11$ and $13$, with
$\chi=16$, which at $d=13$ changes the posterior by at most $7.2\times10^{-10}$ relative to $\chi=32$
(Table~\ref{tab:validation}). On these cells the average is $5.46\pm0.13$ at $d=11$ and $6.86\pm0.27$ at
$d=13$, a step of $1.40\pm0.30$. At $\alpha=0.5$ the $d=13$ value at $p=0.08$, $5.90\pm0.74$, is dominated by one
heavy-tailed seed (single-seed values $4.54$ to $7.09$) and does not resolve a step
(Figure~\ref{fig:scaling}). The uncorrelated controls grow more slowly.
Under X noise the $\alpha$-averaged loss increases from $0.14$ to $1.17$, while the value at
$\alpha=0.5$ lies between $0.31$ and $1.26$ without a monotone trend. Under independent X and Z noise
the $\alpha$-averaged loss is $0.49$, $0.94$, $1.29$ and $1.18$, and the value at $\alpha=0.5$ has a
weighted slope of $0.253\pm0.004$ per unit of $d$ ($\chi^2=9.0$ on 2 degrees of freedom). Over the
sampled range, $d\le11$ and $d=13$ at $p=0.08$, the loss of post-selection on the independent gap thus
grows with the distance under every noise model, several times faster under depolarizing noise than
under the uncorrelated controls. For the optimal rule at $\alpha=0.5$ and $p=0.08$, a fit of
$\ln\LA$ against the logarithm of the unselected ML failure rate over $d=5$ to $13$ gives the exponent
$b_{\mathrm{A}}=3.02\pm0.05$, consistent with $b=3.1(1)$ for the toric code under independent bit- and
phase-flip noise \cite{chen2025}. These data do not determine the corresponding exponent of the gap
rule: its fitted value ranges from $1.1$ to $4.6$ across acceptance fractions and fit windows, with
standard errors of $0.2$ to $0.6$.

For the correlation-aware gap, whose steps up to $d=11$ are given in Section~\ref{sec:ca}, the
$\alpha$-averaged loss at $p=0.08$ on the $40\,000$-shot cells is
$3.21\pm0.15$ at $d=11$ and $3.55\pm0.15$ at $d=13$, so the step from $d=11$ to $13$ is $0.34\pm0.22$,
the smallest of the series and less than two standard deviations from zero. The sampled distances do
not distinguish saturation of this residual from slower growth.

\section{Discussion}
\label{sec:discussion}

For the rotated surface code under depolarizing noise, ranking syndromes by the complementary gap of
independent matching accepts syndromes of high failure probability and rejects syndromes of low failure
probability. At distance 11 and half acceptance, the post-selected failure rate is $310$ times the
optimum at $p=0.08$, $51$ times at $p=0.11$ and about $1300$ times at $p=0.05$, where a recomputation
without truncation of the first $10\,000$ shots reproduces the rates. Averaged over acceptance
fractions, the logarithm of the loss grows with distance up to $d=11$ at every rate and up to $d=13$ at
$p=0.08$, faster than under uncorrelated noise. Ranked by the converged posterior, the matching decision is within a factor $1.012$ of optimal
for $\alpha\le0.5$, so the loss is one of ranking. With the same marginal error rates and the same
two-part rule, uncorrelated X and Z noise gives a factor $5.4$ instead of $310$. That control lies
further from threshold, so this comparison combines the correlation with the distance to threshold.
Ranking by the converged posterior of the independent model (rule D), which removes the gap
approximation and keeps the model, still leaves about two thirds of the logarithmic loss at $p=0.08$.
Reweighting the edges for Y errors removes about half of the logarithmic loss. The complementary gap of
the reweighted matching still leaves a factor of $10$ to $23$ averaged over acceptance fractions at
$d=11$, against about $3$ under uncorrelated noise. This residual grows with distance at $p=0.05$ and
$0.08$, by decreasing steps at $p=0.08$ that reach $0.34\pm0.22$ from $d=11$ to $13$, and is flat between
$d=9$ and $11$ at $p=0.11$.

Since rule D attributes most of the loss to the independent model rather than to the gap, we conjecture
that the mechanism is the weighting of Y errors. Independent matching assigns a Y
error the cost of two independent errors, one in each part, whereas the model assigns it the
probability of a single error. Under this conjecture, competing classes reachable through Y
errors are then much more probable than the two part gaps indicate, to a degree that varies between syndromes with equal $G$.
This is consistent with the wide spread in Figure~\ref{fig:calib}, which is absent under both
uncorrelated controls, and with the limited gain of rule B2, which refines the ordering within a gap
value but still treats the parts as independent.

The comparison is made at code capacity, where the converged posterior is computable for every shot up
to $d=11$. Circuit-level noise also correlates the X and Z parts, through Y errors and two-qubit gate
faults, so the size of a comparable loss there remains open. The residual loss of the reweighted gap
indicates what a soft output must capture under correlated noise: the probability mass of competing
classes reached by changing both parts together, which a gap computed one part at a time does not
price. Joint correlated matching \cite{criger2018,higgott2023belief}, separate treatment of the two
logical operators \cite{seip2026} and learned confidences \cite{dentelski2026} are candidates to
capture it. They define no standard complementary gap, and our conclusion concerns the gaps of
independent and of correlation-aware matching. Converged
class posteriors provide the reference against which such soft outputs can be calibrated.

\section*{Data and code availability}

The decoder, the sampling and analysis scripts, the per-shot data and the logs from which every number
is taken are available from the author on request.

\bibliographystyle{unsrtnat}
\bibliography{refs}

\end{document}